\documentclass[]{raa}
\usepackage{graphicx}
\usepackage{times}
\usepackage{natbib}
\usepackage{epsfig}
\usepackage{longtable,booktabs}
\usepackage{color}
\usepackage{amssymb}
\usepackage{hyperref} 

\newcommand       \mum          {\,{\rm \mu m}}
\newcommand       \ppm          {\,{\rm ppm}}

\newcommand       \simali       {\sim\,}

\begin{document}

\title{Unusual Infrared Emission toward Sgr B2: Possible Planar C$_{24}$}

\author{X. H. Chen\inst{1, 2}\and F. Y. Xiang\inst{1}\and X. J. Yang\inst{1}\and Aigen Li\inst{2}}
\institute{Department of Physics, Xiangtan University, 411105
  Xiangtan, Hunan Province, China; {\sf chenxh@smail.xtu.edu.cn, fyxiang@xtu.edu.cn, xjyang@xtu.edu.cn}\\
\and
Department of Physics and Astronomy, University of Missouri, Columbia, MO 65211, USA; {\sf lia@missouri.edu}\\
}

\abstract{
Interstellar graphene could be present in the interstellar medium (ISM), resulting from the photochemical processing of polycyclic aromatic hydrocarbon (PAH) molecules and/or collisional fragmentation of graphitic particles. Indeed, by comparing the observed ultraviolet (UV) extinction and infrared (IR) emission of the diffuse ISM with that predicted for graphene, as much as $\simali$2\% of the total interstellar carbon could have been locked up in graphene without violating the observational constraints. While the possible detection of planar C$_{24}$, a small piece of a graphene sheet, has been reported towards several Galactic and extragalactic {\it planetary nebulae}, graphene has not yet been detected in {\it interstellar} environments. In this work, we search for the characteristic IR features of C$_{24}$ at $\simali$6.6, 9.8, 20$\mum$ toward Sgr B2, a high-mass star formation region, and find a candidate target toward R.A. (J2000) = $267^{\circ}.05855$ and Decl. (J2000) = $-28^{\circ}.01479$ in Sgr B2 whose {\it Spitzer}/IRS spectra exhibit three bands peaking at $\simali$6.637, 9.853 and 20.050$\mum$ which appear to be coincident with that of C$_{24}$. Possible features of C$_{60}$ are also seen in this region. The candidate region is a warm dust environment heated by massive stars or star clusters, associated with a WISE spot (a tracer of star-formation activity), close to the HII region candidate IRAS 17450-2759, and is surrounded by seven young stellar object candidates within $\simali$5$^{\prime}$, suggesting that the creation and/or excitation of C$_{24}$ could be related to star formation activities.
\keywords{line: identification -- ISM: lines and bands -- ISM: molecules}
}

\authorrunning{Chen et al.}

\titlerunning{Possible Planar C$_{24}$}
\maketitle

\section{Introduction}
Carbon, with many allotropes known to be present in the interstellar medium \cite[ISM; e.g., see][]{2011IAUS..280..416J}, plays an important role in the physical and chemical evolution of the ISM \citep{1998Sci...282.2204H}. Several carbon nanostructures such as PAHs \citep[e.g.,][]{2008ARA&A..46..289T}, nanodiamonds \citep{GLR1999, VTW2002}, fullerenes \citep[e.g.,][]{2010Sci...329.1180C, 2010ApJ...722L..54S, G2010, GRL2011, ZK2011, ZK2013, 2017A&A...605L...1B} and their ions \citep[e.g.,][]{1994Natur.369..296F, 2013A&A...550L...4B, 2015Natur.523..322C, C2016, 2015A&A...584A..55S} are promising carries of many IR emission features seen in interstellar and circumstellar medium, although the exact identification still remains debated \citep[see e.g.,][]{KZ2011, Y2013, R2014, AD2017}.

Graphene was first synthesized in the laboratory in 2004 by A.K.~Geim and K.S.~Novoselov \citep[see][]{NGM2004} for which they received the 2010 Nobel Prize in physics. \citet{2011ApJ...737L..30G, 2012ApJ...760..107G} first detected the unusual IR emission features at $\simali$6.6, 9.8, and 20$\mum$ in several Galactic and extragalactic planetary nebulae (PNe), which are coincident with the strongest transitions of a planar graphene sheet C$_{24}$ theoretically predicted by \citet{KD2011}.\footnote{%
A small planar graphene sheet is essentially a fully dehydrogenated PAH molecule. The planar C$_{24}$ graphene can be considered as coronene whose hydrogen atoms are completely lost, differing from the small fullerene C$_{24}$.  \cite{2017ApJ...836..229B} argued that the 11.3$\mum$ feature, commonly attributed to PAHs, could also arise from fullerene C$_{24}$.}
More recently, \citet{2017ApJ...850..104C} have studied the UV absorption and IR emission of graphene C$_{24}$. They estimated the abundance of graphene in the ISM to be $<$\,5$\ppm$ of C/H (i.e., $\simali$1.9\% of the total interstellar C) by comparing the observed UV extinction and IR emission of the diffuse ISM with that predicted for graphene.

In principle, graphene could be present in the ISM as it could be formed from the photochemical processing of PAHs, which are abundant in the ISM, through a complete loss of their H atoms \citep[e.g., see][]{BT2012}. \citet{CKB2010} showed experimentally that C$_{60}$ could be formed from a graphene sheet. \citet{BT2012} further proposed that such a formation route could occur in space.\footnote{If there exists in the ISM a population of HAC-like nanoparticles with a mixed aromatic/aliphatic structure \citep[e.g., see][]{KZ2011}, a complete loss of their H atoms could also convert HAC-like nanoparticles into graphene \citep[e.g., see][]{G2010, 2011ApJ...737L..30G, GRL2011}. Also, graphene could be generated in the ISM from the exfoliation of graphite as a result of grain-grain collisional fragmentation. It is worth noting that graphite is thought to be a major dust component in the ISM \citep{DL1984} and as mentioned earlier, presolar graphite grains have been identified in primitive meteorites.
}
In these scenarios, the formation of C$_{60}$ from graphene and the formation of graphene from PAHs are more likely to occur in regions rich in energetic UV photons. One such region is Sgr B2, a UV-rich high-mass star formation region. To this end, we search for the IR emission feature of C$_{24}$ towards Sgr B2.

The star formation activity is overall deficient in the Central Molecular Zone (CMZ) of the Galaxy relative to its abundant gas \citep{1983A&A...117..343G, 1996ARA&A..34..645M, 2017A&A...603A..89K}, challenging the empirical rations between star formation rate and gas surface \citep{2018ApJ...853..171G}. The distinctive physical and chemical parameters, such as pressure, temperature, velocity dispersion, abundances of elements, heating/cooling and chemical evolution, in the CMZ \citep[][and references therein]{2006A&A...455..971R, 2012MNRAS.425..720S, 2015MNRAS.450.2094G, 2016A&A...586A..50G, 2016MNRAS.457.2675H, 2018ApJS..236...40T} are therefore of much  interest. Most of the dust ridge clouds in the CMZ contain several thousand $M_{\odot}$ in stars or $< 8\%$ of star-formation efficiency \citep{2017MNRAS.469.2263B}. Despite the active star formation in Sgr B2 \citep[which contains actively forming star clusters, high-mass young stellar objects and many compact H II regions, e.g.,][]{1995ApJ...449..663G, 2015ApJ...815..106H}, the overall cloud appears to be as inefficient as the other dust ridge clouds. We are thus eagerly to investigate the environment in the Sgr B2.

This paper is organized as follows.
A brief description of the {\it Spitzer Space Telescope} spectroscopic data \citep{2004ApJS..154...18H} we used is presented in Section 2, while the results of search are reported and discussed in Section 3. The main conclusion is given in Section 4.

\section{Data Description}\label{section:data}

The data analyzed here were obtained with the {\it Infrared Spectrograph} (IRS) on board the {\it Spitzer Space Telescope} and are publicly available\footnote{Download from \url{http://sha.ipac.caltech.edu/applications/Spitzer/SHA/}.} \citep{2004ApJS..154...18H, 2004ApJS..154....1W}. The data products for the short-low (SL) and long-low (LL) modules from the SSC pipeline (version 18.18) were used to produce the final merged spectra, making use of the four slits: SL2 (5.21 -- 7.56$\mum$) and SL1 (7.57 -- 14.28$\mum$) with resolution $R \sim 100$, and LL2 (14.29 -- 20.66$\mum$) and LL1 (20.67 -- 38.00$\mum$) with resolution $R \sim 100$ \citep[see e.g.,][]{S2007, S2018}. In particular, the SL2, SL1 and LL2 slits cover the wavelength regions of the most pronounced theoretical IR emission features of graphene C$_{24}$ \citep[i.e., at $\simali$6.6, 9.8 and 20$\mum$, see e.g.,][and references therein]{M1996, KD2011, 2017ApJ...850..104C}. We downloaded all the SL and LL data (103 pointings\footnote{We have searched for the IRS Enhanced data towards Sgr B2 within a radius of $\simali$1$^{\circ}$ and obtained 103 pointings in total.}) toward Sgr B2 and used the software MATLAB to extract spectral information from the original data.

\begin{figure}[!ht]
\centering
\includegraphics[width=0.44\textwidth,angle=180]{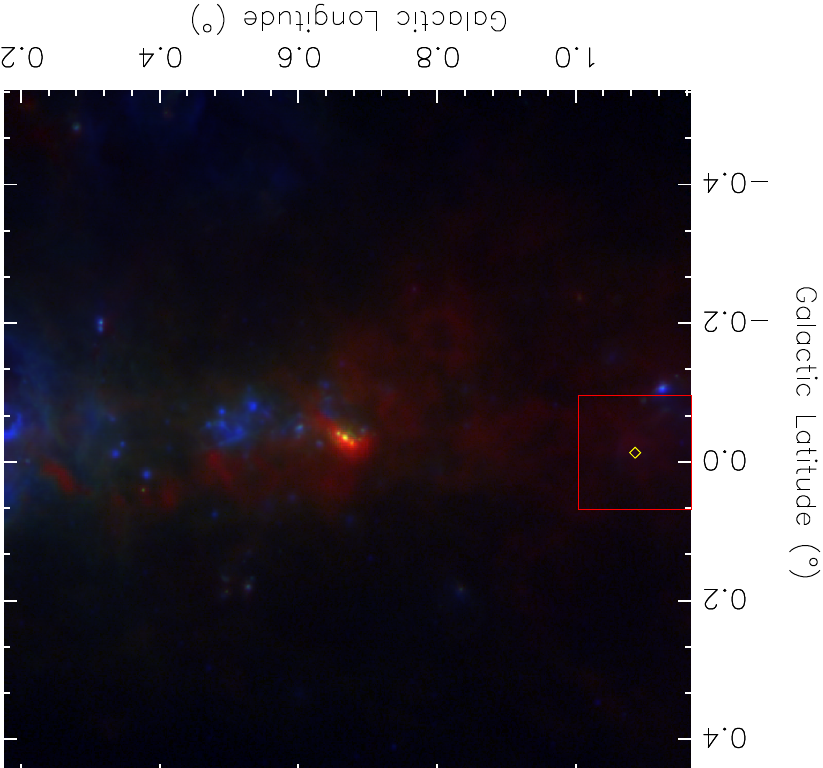}
\includegraphics[width=0.44\textwidth,angle=180]{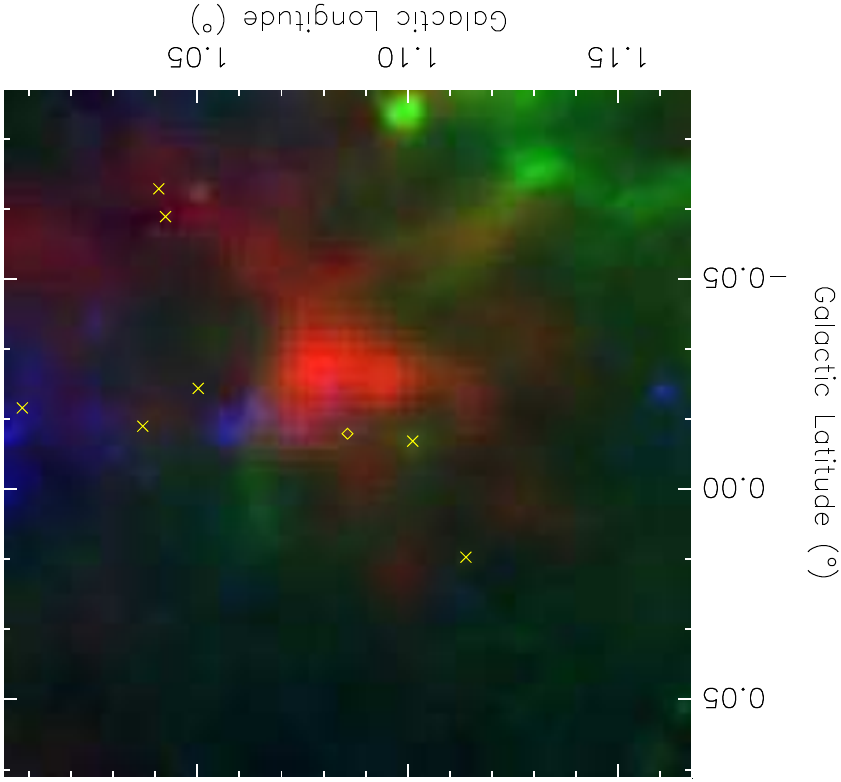}
\caption{Three-color images of the Sgr B2 region, where blue is the 21$\mum$ Band-E MSX image \citep{P2001}, green is the 70$\mum$ image from Herschel InfraRed Galactic Plane Survey \citep[Hi-GAL][]{M2011} which is taken with the Photodetector Array Camera and Spectrometer \citep{PWG2010} on Herschel Space Observatory \citep{PRP2010}, and red is the 500$\mum$ image from Hi-GAL taken with the Spectral and Photometric Imaging Receiver \citep{GAA2010}. The right panel is the close-up image of the red open box in the left panel.The diamond marks the target, and the crosses mark the possible YSOs \citep{2008AJ....136.2413R, 2009ApJ...702..178Y}.}
\label{fig:IRmap}
\end{figure}

We searched for the simultaneous presence of all three emission features of graphene C$_{24}$ at $\simali$6.6, 9.8 and 20$\mum$ and found only four Sgr B2 pointings which exhibit all these features (see Table 1).\footnote{The PI for these {\it Spitzer}/IRS observations was Lee Armus with the Program ID of 1405, 1412, 1413, and 1419 in stare mode. All the four pointings are towards the object HDE216285 entitled by the PI.} Table 1 lists the coordinates, the minimum signal-to-noise ratio (S/N) of the four slits reported from the original downloaded data, and noise level at 1$\sigma$. Because the maximum distance among the four points is $\simali$0.7$^{\prime\prime}$, we took the mean spectrum toward the four pointings and the central coordinate of the new target is R.A. (J2000) = 267$^{\circ}$.05855 and Decl. (J2000) = -28$^{\circ}$.01479. Figure \ref{fig:IRmap} shows the three-color map to locate the target, where the blue, green, and red images are from the 21$\mum$ emission observed by the {\it MidCourse Space Experiment} \citep[MSX;][]{P2001} and the 70 and 500$\mum$ emission from Herschel \citep{M2011}, respectively.

\begin{table}
\centering
\caption{Information for Targets in Sgr B2 \label{table:target}}
\begin{tabular}{cccccc}
\hline\hline
 Index & R.A. (J2000) & Decl. (J2000) & S/N & $\sigma$ \\
       & (degree)     & (degree)      &     & (Jy) \\
\hline
 1    & 267.05851 & -28.01484 & 53 & 0.013 \\
 2    & 267.05847 & -28.01476 & 53 & 0.013 \\
 3    & 267.05857 & -28.01479 & 53 & 0.014 \\
 4    & 267.05866 & -28.01478 & 53 & 0.013 \\
 \hline
\end{tabular}
\end{table}

\section{Results and Discussion}\label{section:result}

\subsection{The Spectra}

\begin{figure*}[!ht]
\centering
\vspace{0.2cm}
\includegraphics[width=0.7\textwidth]{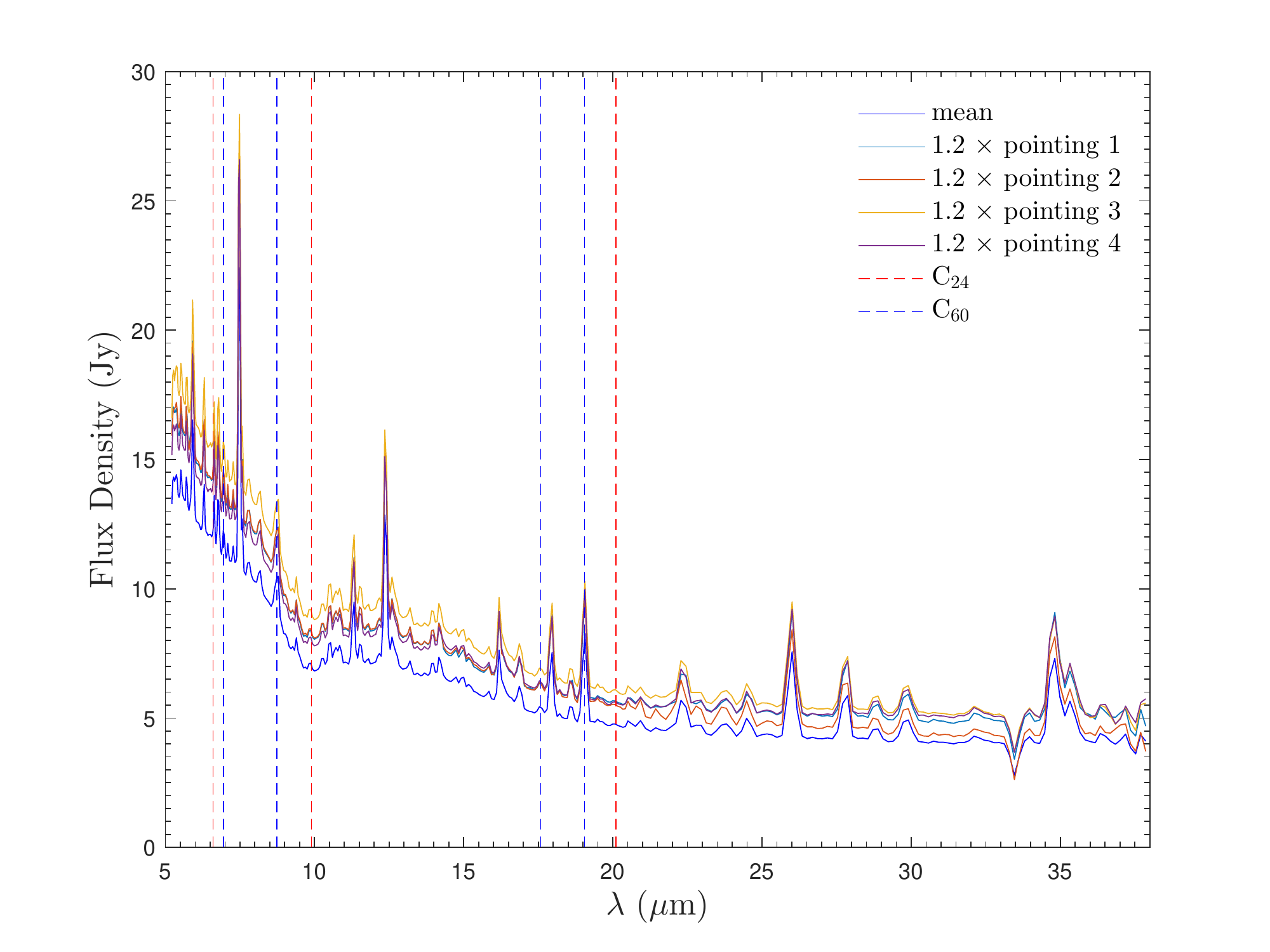}
\caption{The spectra of the four pointings with their flux density multiplied by 1.2 and the mean spectrum of the four pointings. The red and blue dashed lines indicate the possible C$_{24}$ and C$_{60}$ features, respectively.}
\label{Fig:spectra}
\end{figure*}

We detected three IR emission features near the most pronounced theoretical IR emission features of planar C$_{24}$ \citep[i.e., at $\simali$6.6, 9.8 and 20$\mum$, see e.g.,][and references therein]{M1996, KD2011, 2017ApJ...850..104C} towards all the four pointings (see Figure \ref{Fig:spectra}). 
In the following, we shall present the three components based on the mean spectrum in detail.

\begin{figure*}[!ht]
\centering
\vspace{0.2cm}
\includegraphics[width=0.7\textwidth]{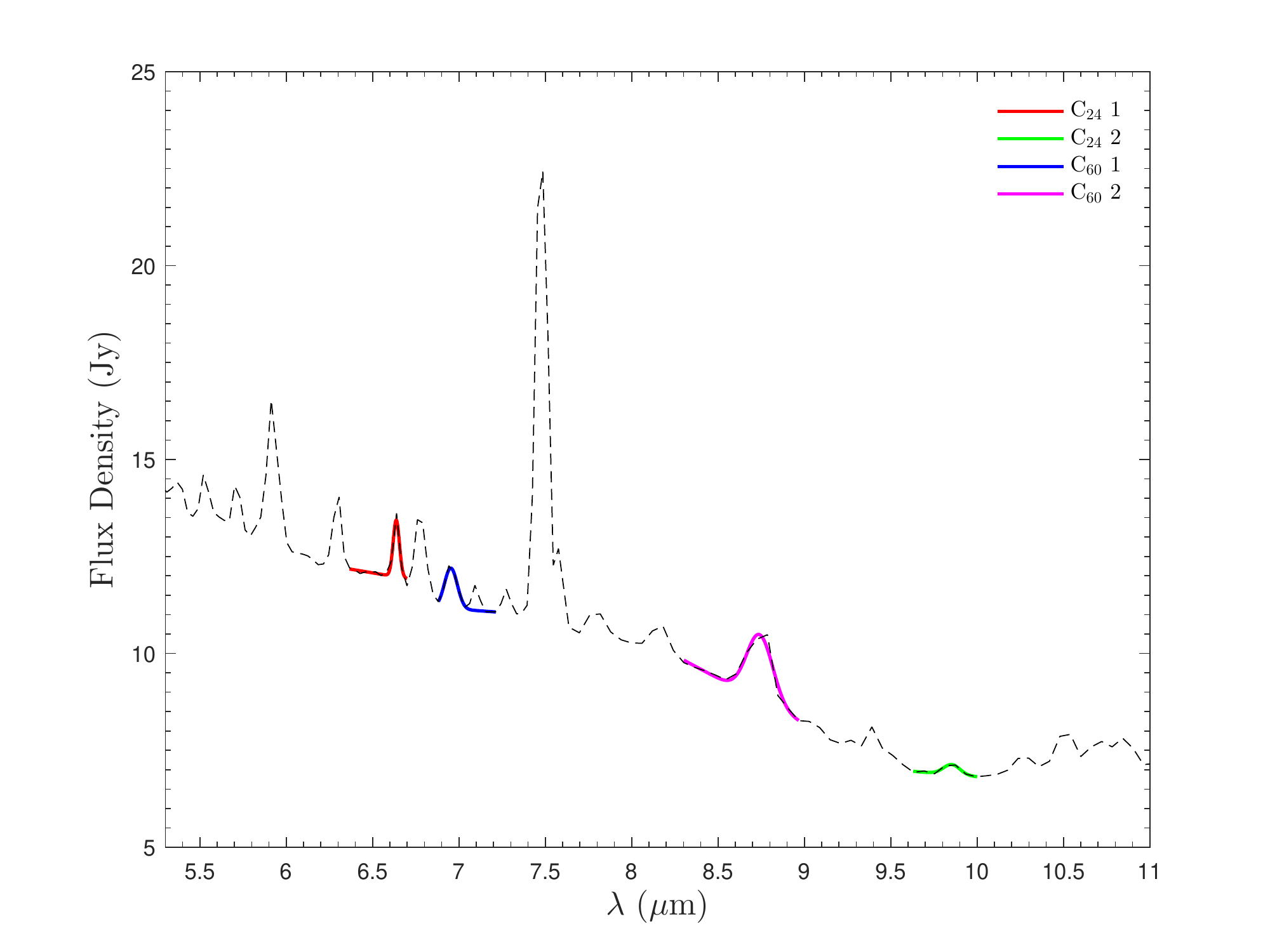}\\
\quad \\
\quad \\
\includegraphics[width=0.7\textwidth]{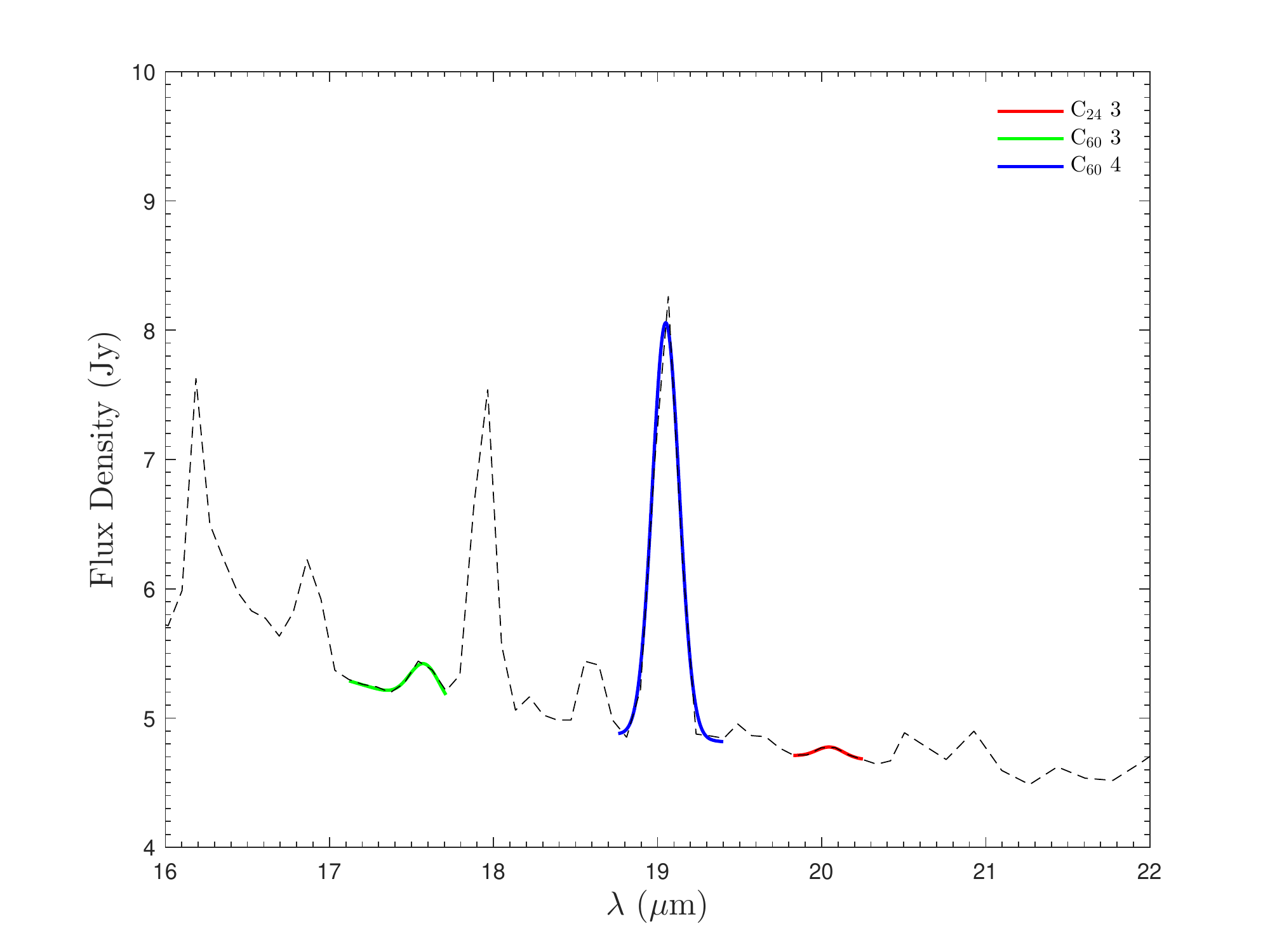}
\caption{The mean spectrum of the four pointings. The color spectra show the results of Gaussian fit for three components of C$_{24}$ and four components of C$_{60}$.}
\label{Fig:spectrum}
\end{figure*}

We fit the spectra in terms of Gaussian profile (see Equation \ref{Equ:fit}) in three windows centered at $\simali$6.64, 9.85 and 20.05$\mum$ for C$_{24}$, where a linear baseline is used in these windows with small wavelength span (i.e., $\lesssim 1$$\mum$):
\begin{equation}\label{Equ:fit}
  F_{\lambda} = A_0 \cdot \exp\left[-\left(\frac{\lambda-\lambda_0}{\Delta \lambda}\right)^2\right]+C_1 \cdot \lambda+C_2,
\end{equation}
where $A_0$, $C_1$ and $C_2$ are constants, $\lambda_0$ and $\Delta \lambda$ are respectively the peak wavelength and the width of the Gaussian profile. We find possible C$_{24}$ emission toward Sgr B2 where the specified coordinates are described in Section \ref{section:data}. The mean spectrum of this candidate is shown in Figure \ref{Fig:spectrum} in detail. The results of the three components are catalogued in Table \ref{table:gaussian fit}. Both $A_0$ of the three components are higher than 5 $\sigma$. The three central wavelengths are all close to those for the most pronounced theoretical IR emission features of planar C$_{24}$ \citep[i.e., at $\simali$6.6, 9.8 and 20$\mum$, see e.g.,][and references therein]{M1996, KD2011, 2017ApJ...850..104C}. That suggests that those three components are likely to originate from the planar, C$_{24}$. The differences of the central wavelengths between the results reported in Table \ref{table:gaussian fit} and theoretical central wavelengths may be due to the well-known fact that DFT computation of C$_{24}$ may not be precise in wavelength \citep{2012JPCA..116.3866B}.

\begin{table}\footnotesize
\centering
\caption{Gaussian Fits to Three Possible Components of C$_{24}$ \label{table:gaussian fit}}
\begin{tabular}{cccccc}
\hline\hline
 Components & $\lambda_0$ & $\Delta \lambda$ & $A_0$ & $C_1$              & $C_2$ \\
          & ($\mu$m)    & ($\mu$m)         & (Jy)  & (Jy$\mum$$^{-1}$) & (Jy) \\
\hline
  1  & 6.637 $\pm$ 0.001  & 0.026 $\pm$ 0.001 & 1.463 $\pm$ 0.025 & -0.747 $\pm$ 0.083 & 16.930 $\pm$ 0.540 \\
  2  & 9.853 $\pm$ 0.001  & 0.067 $\pm$ 0.002 & 0.260 $\pm$ 0.005 & -0.383 $\pm$ 0.022 & 10.650 $\pm$ 0.210 \\
  3  & 20.050 $\pm$ 0.000 & 0.116 $\pm$ 0.002 & 0.082 $\pm$ 0.001 & -0.711 $\pm$ 0.003 & 6.119 $\pm$ 0.066 \\
  \hline
\end{tabular}
\end{table}

The integrated intensity ($I$) ratios for C$_{24}$ calculated from the fitted line are $I_1/I_2 \approx 2.2$ and $I_1/I_3 \approx 4.0$, where subscripts denote component indexes. The observed $I_1/I_2$ here is comparable to the theoretical value of $\simali$1.8, but the observed $I_1/I_3$ is higher than the theoretical value of $\simali$0.4 predicted for C$_{24}$ excited by the interstellar radiation field of the general diffuse ISM \citep{2017ApJ...850..104C}. The band ratios are sensitive to the starlight spectral shape (i.e., ``hardness'') and, to a less degree, to the starlight intensity \citep[see][]{2001ApJ...551..807D}. An exact match would require detailed IR emission modeling of C$_{24}$ in Sgr B2 in the future.

Because the C$_{24}$ features have been detected in conjunction with fullerene features in Galactic and extra-galactic PNes \citep{2011ApJ...737L..30G, 2012ApJ...760..107G}, 
we also fit the spectrum by Gaussian profiles (see Equation \ref{Equ:fit}) in four windows centered at $\simali$7.0, 8.8, 17.4 and 18.9$\mum$ for C$_{60}$ (see Figure \ref{Fig:spectrum}).
The results are cataloged in Table \ref{table:gaussian fit2}.
The C$_{60}$ integrated intensity ratios are $I_1/I_4\sim 0.16$, $I_2/I_4 \sim 0.52$, and $I_3/I_4 \sim 0.11$. These band ratios differ from those of PNe \citep{2010ApJ...722L..54S} and reflection nebulae \citep{2012ApJ...757...41B}. But this could be simply due to the fact that the excitation condition of C$_{60}$ varies from our target to those of PNe and reflection nebulae \citep{2011MNRAS.413..213I}. Both the intensity and ``hardness'' of the starlight radiation fields to which C$_{60}$ is exposed would affect the temperature distribution of C$_{60}$. Note that C$_{60}$, like the C$_{24}$ graphene, undergoes single-photon heating and the maximum temperature it attains is sensitive to the spectral shape of the illuminating starlight \citep[see][]{2001ApJ...551..807D}. 

The coexistence of C$_{24}$ and C$_{60}$ further confirms that the IR emission at $\simali$6.637, 9.853 and 20.050$\mum$ may originate from planar C$_{24}$. \citet{2011ApJ...737L..30G} propose that shocks which are driven by strong stellar winds, can trigger HACs' processing. Fullerenes,  possibly planar C$_{24}$ molecules, and other complex aromatic and aliphatic species, could evolve from the vaporization of HACs.

\begin{table}\footnotesize
\centering
\caption{Gaussian Fits to Four Possible Components of C$_{60}$ \label{table:gaussian fit2}}
\begin{tabular}{cccccc}
\hline\hline
 Components & $\lambda_0$ & $\Delta \lambda$ & $A_0$ & $C_1$              & $C_2$ \\
          & ($\mu$m)    & ($\mu$m)         & (Jy)  & (Jy$\mum$$^{-1}$) & (Jy) \\
\hline
  1  & 6.953 $\pm$ 0.001   & 0.053 $\pm$ 0.001 & 1.047 $\pm$ 0.022 & -0.321 $\pm$ 0.102 & 13.380 $\pm$ 0.730 \\
  2  & 8.742 $\pm$ 0.001   & 0.107 $\pm$ 0.002 & 1.696 $\pm$ 0.024 & -2.364 $\pm$ 0.056 & 29.450 $\pm$ 0.480 \\
  3  & 17.580 $\pm$ 0.000  & 0.130 $\pm$ 0.003 & 0.304 $\pm$ 0.008 & -0.373 $\pm$ 0.023 & 11.680 $\pm$ 0.390 \\
  4  & 19.050 $\pm$ 0.000  & 0.115 $\pm$ 0.001 & 3.029 $\pm$ 0.011 & -0.090 $\pm$ 0.019 & 6.564 $\pm$ 0.354 \\
  \hline
\end{tabular}
\end{table}

As discussed above, we refer to the candidate carrier of the three components (centered at at $\simali$6.637, 9.853 and 20.050$\mum$) as planar C$_{24}$. Further observations, especially with higher spectral resolution, and further experimental research and theoretical computation are required to confirm this.

\subsection{Relation with Star-Formation Activity}

The excitation of C$_{24}$ requires UV photons. As illustrated in Figure~2 of \citet{2017ApJ...850..104C}, C$_{24}$ mostly absorbs in the far-UV. Therefore, in regions with intense star-formation activities, the excitation of C$_{24}$ is naturally expected to occur. Also, the creation of C$_{24}$ from PAHs, HAC  and/or graphite is also more likely to occur in UV-rich regions.

\begin{figure*}[!ht]
\centering
\includegraphics[width=0.7\textwidth]{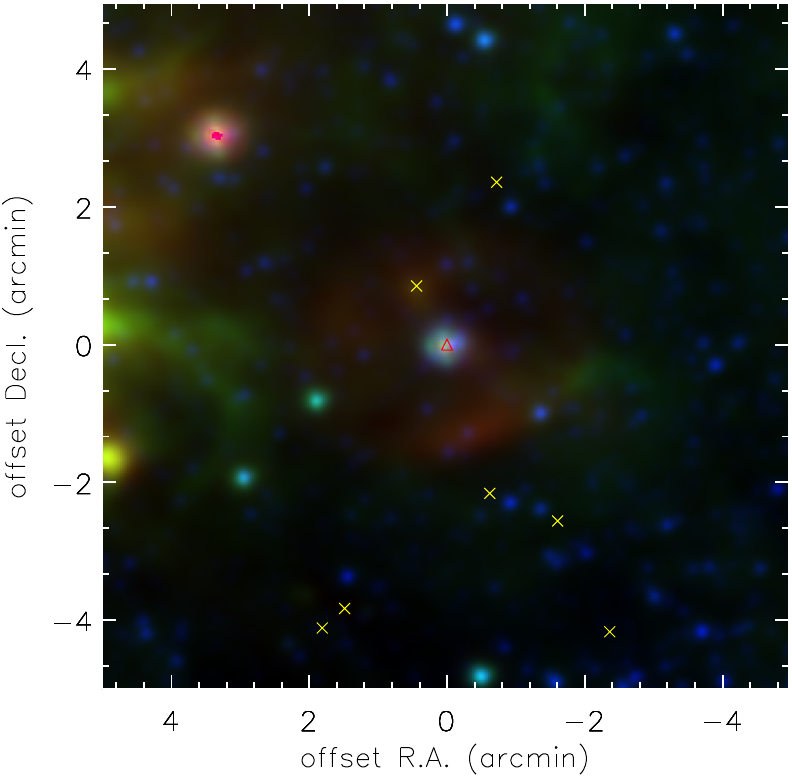}
\caption{Background false-color map centered on the target (red triangle) at R.A. (J2000) = 267$^{\circ}$.05855 and Decl. (J2000) = -28$^{\circ}$.01479) with seven YSO candidates superimposed \citep[yellow crosses,][]{2008AJ....136.2413R, 2009ApJ...702..178Y}, where blue is the WISE 4.6$\mum$, green is 12$\mum$, and red is 22$\mum$ data.}
\label{Fig:wise}
\end{figure*}

In addition to the 21$\mum$, 70 and 500$\mum$ images \citep[see Figure \ref{fig:IRmap},][]{P2001, M2011}, we also provide high-sensitivity mid-infrared images from the Wide-field Infrared Survey Explorer \citep[WISE, see Figure \ref{Fig:wise},][]{2010AJ....140.1868W} to indicate the protostar activity. The candidate YSOs are superposed in these two figures. The 21 and 70$\mum$ emission show the warm dust heated by nearby massive stars or star clusters; and the far-infrared emission (e.g. at 500$\mum$) that is from the cold dust, shows the locations of the dense molecular clouds that have few indicators of active star formation \citep[][and references therein]{S2018}. As stated in \citet{2010AJ....140.1868W}, the excesses at 12$\mum$ and 22$\mum$ bands are indicator of star formation activity. Both Figures \ref{fig:IRmap} and \ref{Fig:wise} indicate that the target is probably impacted by star formation. The former one delineates a star-forming region where the warm dust is heated by nearby massive stars or star clusters and is surrounded by cold dust. The excesses at 12$\mum$ and 22$\mum$ bands in the latter one also imply that the source is probably impacted by star formation.

The target region where C$_{24}$ is seen is more likely associated with HII regions than PNe. The IRAS source (with an angular resolution of $\simali$0.75$'$ -- 3.0$'$), IRAS 17450-2759, of 13.7$''$ away from the target shows $F_{12}/F_{25}= 0.64$, $F_{25}/F_{60} = 0.06$, $F_{60}/F_{12} = 27.58$, $F_{100}/F_{12} = 112.74$, $F_{100}/F_{25} = 71.94$ and $F_{100}/F_{60} = 4.09$, where $F_{12}$, $F_{25}$, $F_{60}$, and $F_{100}$ denote the flux densities at 12, 25, 60 and 100$\mum$. The corresponding flux qualities are $Q_{12} = 3$, $Q_{25} = 2$, $Q_{60} = 1$, and $Q_{100}=3$,\footnote{The flux quality values of 1, 2, and 3 represent an upper limit, moderate quality, and high quality, respectively (see \url{https://heasarc.gsfc.nasa.gov/W3Browse/all/iraspsc.html}).} respectively. Both of them are conformity with criteria of HII regions which are excited by embedded high mass stars \citep[except $Q_{60} > 1$ required in some criteria, see][]{YZW2018} rather than PNe's property \citep{2001A&A...378..843K}.

We have also searched for YSOs from SimBad\footnote{\url{http://simbad.u-strasbg.fr/simbad/sim-fcoo}} within 5$'$ to the target in this work. Seven YSO candidates\footnote{Include SSTGC 891214, SSTGC 899543, SSTGC 909173, 2MASS J17480460-2805032, 2MASS J17480762-2803267 2MASS J17481157-2803027 from \citet{2009ApJ...702..178Y}, and SSTGLMC G001.0423-00.0650 from \citet{2008AJ....136.2413R}.} \citep{2008AJ....136.2413R, 2009ApJ...702..178Y} are surrounding the target, indicating that the candidate planar C$_{24}$ emission is likely to associate with star formation activity.

In addition, the 4.6, 12 and 22$\mum$ filters from WISE data include the continuum emission from ultrasmall grains which undergo stochastic heating \citep{2010AJ....140.1868W}. These nano-sized ultrasmall grains, if they are carbonaneous in nature like HAC, their collisional fragmentation and full dehydrogenation, triggered by UV photons and/or shockwaves from star formation activities, could result in the creation of C$_{24}$. Therefore, it is not surprising that C$_{24}$ is related to star formation activity.

Finally, we note that it would potentially be useful to compare the physical and chemical conditions (e.g., the UV starlight intensities, the hydrogen and electron number densities, and the gas temperature) of the four pointings in which C$_{24}$ is possibly present with that of the other pointings in which C$_{24}$ is not seen. However, it is difficult to obtain a complete census of the YSOs and their initial mass functions in these regions. Also, the UV extinction prevents us from an accurate direct measure of the UV starlight intensities even if the UV photometric or imaging data are available (e.g., from HST). Therefore, we prefer not to make comparisons between the environments of the four pointings and the other pointings.

\section{Summary}

PAHs, one of graphene precursor, are widespread in star-forming environments. Inspired by the possible detection of the C$_{24}$ emission features at $\simali$6.6, 9.8, 20$\mum$ in several Galactic and extragalactic PNe, and theoretical endeavor recently about C$_{24}$ in ISM, we have searched for the characteristic IR emission features of C$_{24}$ toward the high-mass star formation region Sgr B2. We detected IR emission, with three peak wavelengths of $\simali$6.637, 9.853 and 20.050$\mum$ toward R.A. (J2000) = 267$^{\circ}$.05855 and Decl. (J2000) = -28$^{\circ}$.01479 in Sgr B2. These three wavelengths are all very close to the most pronounced three characteristic IR emission features of C$_{24}$. These detected features are also companied with the characteristic IR emission bands of possibly C$_{60}$.

The three-color (21, 70 and 500$\mum$) images indicate that the target is probably in the warm dust environment which is heated by nearby massive stars or star clusters. The WISE false-color map suggests that the target source is associated with a WISE spot with excesses at 12$\mum$ and 22$\mum$, a tracer of star-formation activity. The nearest IRAS source (IRAS 17450-2759) is 13.7$^{\prime\prime}$ away, and this IRAS source is a HII region candidate. The target source is also surrounded by seven YSO candidates within 5$^{\prime}$. All these suggest that the IR emission of C$_{24}$ could be powered by star formation activity.

\section{Acknowledgments}
We thank the anonymous referee for his/her very helpful suggestions and comments which substantially improved the quality of this work. This research is supported by Hunan Provincial Innovation Foundation for Postgraduates CX2015B213, and by the Joint Research Funds in Astronomy (U1531108, U1731106 and U1731110) under cooperative agreement between the National Natural Science Foundation of China and Chinese Academy of Sciences, and in part by NSFC U1731107.

\end{document}